\pdfoutput=1
\documentclass[11pt]{article}

\usepackage{ACL2023}

\usepackage{times}
\usepackage{latexsym}

\usepackage[T1]{fontenc}
\usepackage[utf8]{inputenc}

\usepackage{microtype}

\usepackage{inconsolata}

\usepackage{graphicx} % DO NOT CHANGE THIS
\usepackage{booktabs}
\usepackage{amsthm}
\usepackage{amsmath}
\usepackage{amsfonts}
\usepackage{color}
\usepackage{multirow}
\usepackage{array}
\usepackage{rotating}

\newcolumntype{C}[1]{>{\centering\let\newline\\\arraybackslash\hspace{0pt}}m{#1}}

\usepackage{tcolorbox}
\usepackage{pifont}
\newtcolorbox[list inside=prompt,auto counter,number within=section]{prompt}[1][]{
    colbacktitle=black!60,
    coltitle=white,
    fontupper=\footnotesize,
    boxsep=5pt,
    left=0pt,
    right=0pt,
    top=0pt,
    bottom=0pt,
    boxrule=1pt,
    #1,
}

\title{A Multi-Task Embedder For Retrieval Augmented LLMs}
\author{Peitian Zhang$^{2}$\thanks{~~Co-first authors. This work was done when Peitian Zhang is in an internship at BAAI.}, \ \  \textbf{Shitao Xiao}$^{1*}$, \ \ \textbf{Zheng Liu}$^{1*}$\thanks{~~Corresponding author.}, \ \  \textbf{Zhicheng Dou}$^{2}$, \ \  \textbf{Jian-Yun Nie}$^{1,3}$, 
\\
$^1$Beijing Academy of Artificial Intelligence, \ \ 
$^2$Renmin University of China,  \ \
$^3$University of Montreal 
\\
{\tt \{namespace.pt, zhengliu1026\}@gmail.com }
}

\begin{document}

\maketitle

\begin{abstract}
LLMs confront inherent limitations in terms of its knowledge, memory, and action. The retrieval augmentation stands as a vital mechanism to address these limitations, which brings in useful information from external sources to augment the LLM. 
However, existing retrieval methods encounter two pressing issues. On one hand, the general retrievers are not properly optimized for retrieval augmentation hence exhibit limited effectiveness; on the other hand, the task-specific retrievers excel in the targeted retrieval augmentation scenario, while lack the versatility to handle diverse scenarios. 
In this work, we propose \textbf{LLM-Embedder} for the unified support of diverse retrieval augmentation scenarios. Our method presents three technical contributions. Firstly, we introduce a new \textit{reward formulation}, namely {rank-aware reward}. It exploits the ranking position of the desired output among $N$ sampled outputs from the LLM, which leads to fine-grained and robust computation of reward from the LLM's feedback. Secondly, we design a novel \textit{distillation objective}, called graded distillation. It incorporates both the absolute value and the relative order of the reward for more sufficient utilization of the LLM's feedback. Thirdly, we systematically optimize the \textit{multi-task learning}, which effectively unifies the multiple retrieval functionalities into one model. In our experiment, LLM-Embedder notably improves the LLM's performances in various downstream tasks, and outperforms both general and task-specific retrievers with a substantial advantage.

% while introducing superior retrieval augmentation's effect  over both general and task-specifc retrievers. 

\end{abstract}

% LLM有三方面局限，包括知识，记忆，能力。
% 知识：参数有限，世界知识庞大且时刻变化
% 记忆：context length有限，难以deal with long context
% 能力：需要prompting和example来产生aligned response；同时需要借助工具和外界交互
% RAG是可以有效解决三个局限，介绍RAG，其通过knowledge/memory/example/tool retrieval来enhance llm
% embedding model是RAG核心
% 现有的embedding model两大问题
% 我们提出LLM-Embedder
% 训练这样一个模型很有挑战。
% 总的来说，RAG的愿景是检索回来的内容能够增加模型生成desired output的概率。
% 则第一步是分辨真正有用的文档。传统方法使用likelihood，然而忽视了别的生成结果
% 我们使用relative likelihood reward衡量文档的有效性，其计算正确答案在众多采样结果中的提升排名数after condition on the candidate.

% 基于reward distribution，传统方法使用KL进行知识蒸馏，然而其不能使用in-batch negatives，同时难以处理一些极端的reward distribution，比如polarized and flat。
% 我们提出了graded distillation，其不仅能够incorporate in-batch negatives，同时能够robustly learn from different distributions
% 最后，不同RAG task要求embedder capture不同的semantic relationships，彼此之间可能存在冲突。我们device 3-stage multi-task learning framework to harmonize the optimization process.

\begin{figure}[t]
\centering
\includegraphics[width=\linewidth]{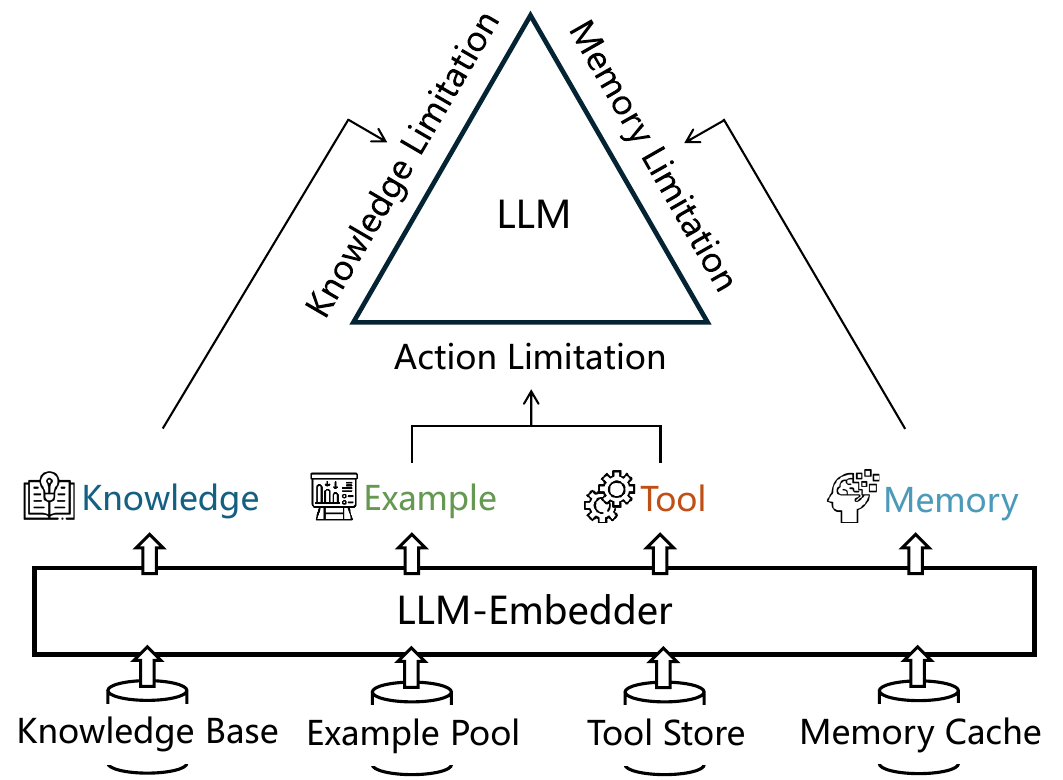}
\vspace{-20pt}
\caption{LLM-Embedder presents a unified embedding model for the diverse retrieval augmentation scenarios.} 
\vspace{-15pt}
\label{fig:llm-embedder}
\end{figure} 

\vspace{-5pt}
\section{Introduction}
Large language models (LLMs) present a unified foundation to support general artificial intelligence applications~\cite{brown2020language,chowdhery2022palm,touvron2023llama}. Despite the substantial improvement over the last-gen methods, LLMs still face many severe problems, such as hallucination~\cite{ji2023survey,bang2023multitask}, limited memory~\cite{bai2023longbench,an2023eval}, mis-following of instructions~\cite{ouyang2022training,bai2022constitutional}. Many of the challenges can be traced back to the inherent limitations of LLMs in terms of \textit{knowledge}, \textit{memory}, and \textit{action}. 
Specifically, LLMs cannot internalize the vast and constantly changed world knowledge due to their finite and static parameters.
LLMs are incapable of memorizing and utilizing long-term information because of the limited context length. 
Finally, LLMs require manually in-context examples and tools to accomplish complex real-world tasks.

Retrieval augmentation stands as a vital mechanism to address these inherent limitations of the LLM. It brings in useful information from external sources, such as knowledge, memory pieces, in-context examples, and tools, which substantially enhances the LLM for the generation of desired outputs~\cite{gao2023retrieval}. 
The embedding model (a.k.a. \textit{embedder}) is a critical part of retrieval augmentation, which bridges the LLM's information needs with external sources. The existing embedding models can be briefly partitioned into two categories. One is the general-purpose embedders, which aim to be universally applicable for various retrieval tasks~\cite{izacard2021unsupervised,wang2022text,xiao2023bge}. Despite their popularity, they are not properly optimized for retrieval augmentation, and are thus prone to an inferior effectiveness in the corresponding task. The other one is the task-specific embedders, which are tailored for one specific retrieval augmentation scenario, e.g., knowledge retrieval~\cite{yu2023augmentation} and example retrieval~\cite{wang2023learning}. However, these methods lack versatility across different scenarios. As the LLMs require assistance from diverse external sources in solving real-world problems, it becomes imperative to develop an effective and versatile embedding model to support the diverse retrieval augmentation needs. 

In this paper, we present LLM-Embedder, a unified embedding model to support a broad range of retrieval augmentation scenarios, including knowledge retrieval, memory retrieval, example retrieval, and tool retrieval. Training such a versatile embedding model presents multiple challenges in terms of 1) how to learn from the LLM, and 2) how to harmonize different retrieval tasks. In LLM-Embedder, the following technical contributions are presented. 

\noindent$\bullet~$\textbf{Reward Formulation.} For each retrieval augmentation scenario, the embedder is learned from the LLM's feedback, i.e. the retrieval candidate needs to be promoted if it contributes to the generation of the desired output. Conventional methods rely on the generation likelihood ~\cite{shi2023replug,atlas2023izacard}. 
% However, the absolute generation likelihood tends to fluctuate dramatically given different training samples, which leads to a poor discriminability of the effect resulted from retrieval. 
However, the absolute generation likelihood tends to fluctuate dramatically, which may lead to inaccurate estimation of the contribution of each retrieval candidate. 
In LLM-Embedder, we propose a new reward formulation called \textit{rank-aware reward}. Essentially, a retrieval candidate will receive a higher reward if it can better promote the desired output's ranking among $N$ sampled outputs from the LLM. Thus, it is free from dealing with the absolute generation likelihood, which facilitates a fine-grained and more robust computation of the reward.  

% \noindent$\bullet~~$\textbf{Reward from the LLM.} Adapting the embedder for one retrieval augmentation scenario needs to evaluate the reward of each retrieval candidate based on the feedback from the LLM.
% Existing methods usually rely on the generation likelihood of the desired output~\cite{shi2023replug,atlas2023izacard}.
% % However, this measurement may be inaccurate if the retrieval candidate pushes the likelihood of the incorrect output even higher.
% However, the absolute likelihood value is not proportional to the generation probability since other generation options may have higher likelihood.
% Therefore, we propose the \textit{rank-aware reward}, which computes the relative rank (instead of absolute likelihood) of the desired output among multiple sampled outputs.
% In this way, we're able to identify the retrieval candidates that truly contribute to the generation of the desired output.

\noindent$\bullet$ \textbf{Distillation Objective}. Based on the LLM's reward, the embedding model is learned by knowledge distillation. Typically, this is accomplished by minimizing the KL-divergence between the reward distributions and the relevance distribution estimated by the embedder~\cite{shi2023replug,yu2023augmentation}. In many cases, the reward distribution are either polarized (extremely high rewards for one candidate while low rewards for others) or flat (even rewards for every candidate), which makes it difficult to distill fine-grained knowledge with KL-Divergence. To address this problem, we design the \textit{graded distillation}. It integrates both the absolute values of rewards and their relative orders for knowledge distillation, which leads to a more sufficient exploitation of the LLM's feedback. 

% \noindent$\bullet~~$\textbf{Learning from the reward distribution.}
% KL-divergence is the most common loss to learn the embedder given the reward distribution~\cite{shi2023replug,yu2023augmentation,atlas2023izacard}. Nevertheless, when the reward distribution is polarized (high reward for one candidate while low for others) or flat (even reward for each candidate), KL-divergence becomes ineffective.
% To address this problem, we innovate a \textit{graded distillation} objective, which integrates a series of contrastive losses whose labels are transformed from the rewards.
% It can robustly optimize the embedder from various reward distributions and incorporates in-batch negatives to further improve the embedder's discrimination ability. 

\noindent$\bullet~$\textbf{Multi-task Learning.} LLM-Embedder is trained to support diverse retrieval augmentation scenarios through multi-task learning. However, different scenarios need to capture distinct semantic relationships, hence the multiple training tasks may conflict with each other. To harmonize the learning process, we perform systematic optimization with three techniques: 1) \textit{self-paced learning scheduling}, where lossy tasks can be automatically compensated by higher learning rates; 2) \textit{homogeneous batching}, where training samples from one common task are gathered in the same batch to optimize the impact of in-batch negative sampling; 3) \textit{diversified prompting}, which presents different tasks with unique prefixes such that the embedding model can better distinguish each of them. 

To summarize, LLM-Embedder stands as a pioneering work for the uniform support of the diverse retrieval augmentation scenarios of LLMs. It makes threefold technical contributions, and brings valuable inspirations on how to learn from LLM's feedback and how to harmonize different retrieval tasks. In our experiment, LLM-Embedder achieves a superior performance, where it notably improves the LLM's performance in a variety of downstream tasks. Meanwhile, its retrieval augmentation's effect is superior to both general and task-specific retrieval methods. Our model and code will be publicly available to facilitate future research. 

% versatile method to support the diverse retrieval augmentation scenarios of LLMs. It makes threefold technical contributions, and brings valuable inspiration on how to learn from LLM's feedback and how to coordinate the different retrieval tasks. In our experiment, LLM-Embedder achieves a superior performance, where it notably improves the LLM's performance in a variety of downstream tasks. Meanwhile, its retrieval augmentation's effect is superior to both general and task-specific retrieval methods. Our model and code will be publicly available to facilitate future research. 

% To summarize, LLM-Embedder stands as a versatile method to support the diverse retrieval augmentation scenarios of LLMs. It makes threefold technical contributions, and brings valuable inspiration on how to learn from LLM's feedback and how to coordinate the different retrieval tasks. In our experiment, LLM-Embedder achieves a superior performance, where it notably improves the LLM's performance in a variety of downstream tasks. Meanwhile, its retrieval augmentation's effect is superior to both general and task-specific retrieval methods. Our model and code will be publicly available to facilitate future research. 

\section{Related Works}
\noindent$\bullet~~$\textbf{Embedding Model} maps the input text into dense vector (i.e. \textit{embedding}) in the semantic space, where the relevance between texts is measured by the similarity between embeddings.
It has become the de-facto choice for modern information retrieval systems.
There are mainly three research threads for improving the performance of embedding models.
The first one is leveraging advanced backbone models, including the retrieval oriented models~\cite{liu2022retromae,wang2022simlm} and large language models~\cite{repllama2023ma,llara2024li}.
Another thread is enhancing the learning methodology, such as upgrading the negative sampling strategy~\cite{karpukhin2020dense,izacard2021unsupervised,xiong2020approximate} and incorporating knowledge distillation from a more precise ranking model~\cite{qu2020rocketqa,hofstatter2021efficiently,xiao2022distill}.
Last but not least, many recent works dedicate to train a universal retriever across a wide array of tasks~\cite{wang2021gpl,lewis2021paq,karouzos2021udalm,yu2022coco,su2022one,asai2022task}.
LLM-Embedder inherits successful practices for training high-quality dense retriever, while innovating novel techniques to tailor for the multi-task learning of diverse retrieval augmentation scenarios. 

\noindent$\bullet~~$\textbf{Retrieval Augmentation} is a vital mechanism to address the inherent limitations of the LLM in terms of knowledge, memory, and action. Concretely, the LLM can 1) generate factoid answers with retrieved knowledge~\cite{rag_survey2024gao,jiang2023active}; 2) utilize long-context information with retrieved memory pieces~\cite{rubin2023long,wang2023augmenting,retrieval_meets_long_context2023xu}; 3) better follow human instruction with retrieved in-context examples~\cite{gpt3_2020brown,cheng2023uprise}; 4) execute complex tasks with retrieved tools~\cite{qin2023toolllm}.
In practice, there are two common options of retrievers: the general retrievers~\cite{robertson2009probabilistic,izacard2021unsupervised,xiao2023bge,neelakantan2022text} and the task-specific retrievers~\cite{yu2023augmentation,wang2023augmenting,qin2023toolllm}.
The general retrievers exhibit superior versatility, but may suffer from an inferior retrieval quality in retrieval augmentation tasks. 
In contrast, task-specific retrievers are more specialized, achieving better performance in the targeted scenario, while falling short when handling other scenarios. 
Compared with the existing works, LLM-Embedder unifies the generality and specialty: it comprehensively supports all major retrieval augmentation needs of the LLM, meanwhile achieving the leading performance in every retrieval augmentation scenario. 

\section{LLM-Embedder}

\begin{figure}[tb]
    \centering
    \includegraphics[width=\linewidth]{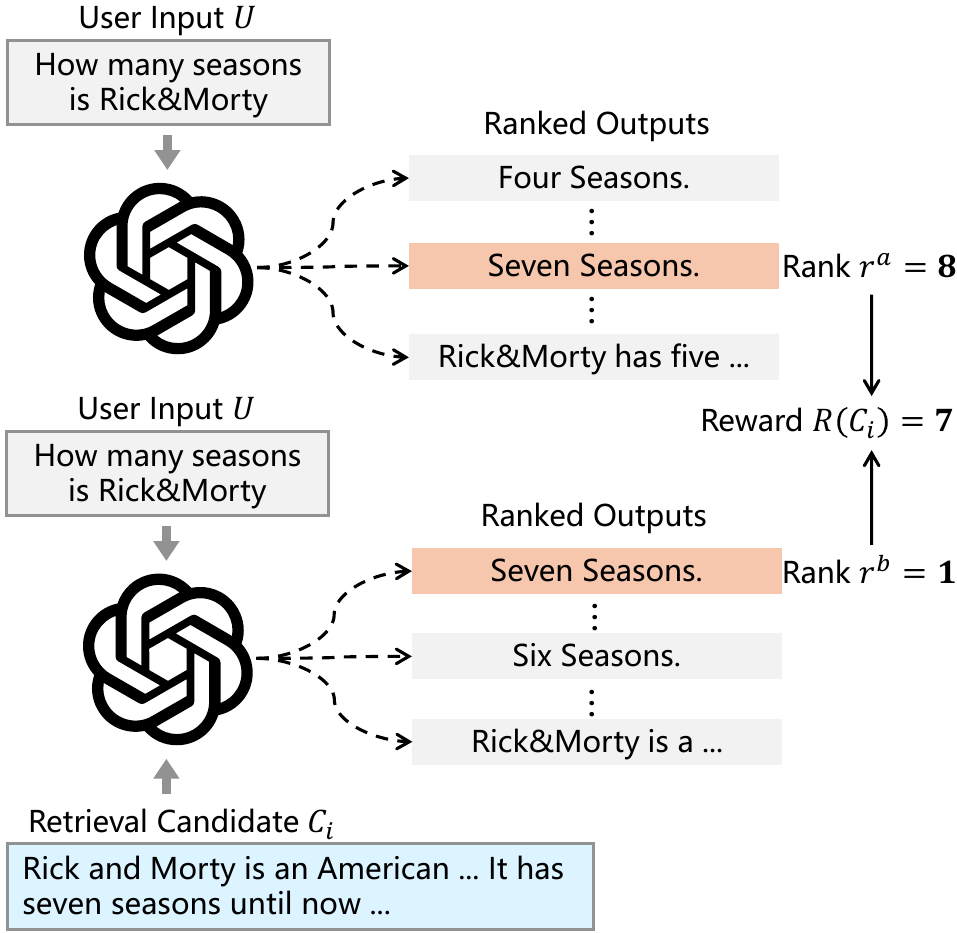}
    \vspace{-20pt}
    \caption{The rank-aware reward for each retrieval candidate. It measures the improvement of the rank of the desired output among multiple sampled outputs.}
    \label{fig:reward}
    \vspace{-10pt}
\end{figure}

In this section, we will present the retrieval augmentation scenarios with LLM-Embedder ($\S$\ref{sec:method-rag}), and introduce its training methodology ($\S$\ref{sec:method-train}). 

\subsection{Retrieval Augmentation}\label{sec:method-rag}
% \subsection{Retrieval Augmentation of the LLM}\label{sec:method-rag}
% LLM-Embedder支持LLM四种主要的Retrieval augmentation needs of the LLM
% 介绍检索 -> formalize检索 -> formalize rag -> 介绍四个场景
% 

LLM-Embedder targets on the unified support for the major retrieval augmentation needs of the LLMs, including knowledge retrieval, memory retrieval, example retrieval, and tool retrieval.
It transforms each retrieval candidate $C_i\in\mathcal{C}$ into its embedding $\boldsymbol{C}_i\in\mathbb{R}^D$ and stores all embeddings in a vector DB. It also embeds the user input $U$ into $\boldsymbol{U}\in\mathbb{R}^D$, then retrieves the top-$K$ relevant candidates based on cosine similarity:
\begin{equation}
    \mathrm{Ret}(U)\gets\underset{C_i}{\mathrm{top\text{-}}K}\{\mathrm{cos}(\boldsymbol{U},\boldsymbol{C}_i)\}.
\end{equation}
The retrieval result and the user input are synthesized with template $\psi$ to prompt the LLM $\Theta$:
\begin{gather}
    O\gets \Theta(\psi(U,\mathrm{Ret}(\boldsymbol{U}))).
\end{gather}
Each retrieval augmentation scenario has its unique formulation of retrieval candidate, user input, and prompt template, which are elaborated as follows. 

\noindent$\bullet~~$\textbf{Knowledge Retrieval}.
The LLM can generate factoid answers with retrieved knowledge.
Each retrieval candidate is a passage from an external knowledge corpus.
The user input is usually an explicit question.
It can also be a conversation context with a context-dependent question.
In this case, we concatenate the entire context as the user input.
The retrieved passages and the user input are synthesized according to Template~\ref{prompt:knowledge}.

\noindent$\bullet~~$\textbf{Memory Retrieval}. 
The LLM can remember and utilize long context memory with memory retrieval~\cite{retrieval_meets_long_context2023xu}.
Specifically, the long context split into equal-size chunks $\{v_1,\dots,v_n\}$. When processing the $v_j$, each previous chunk concatenated with its subsequent chunk is treated as a retrieval candidate, i.e. $C_i\gets v_i+v_{i+1},~~i<j$. The user input is $v_i$ itself.
Denote the LLM's context window size as $L^*$. We maintain the recent $L$ tokens in the context window, while the rest $L^*-L$ are populated with retrieved chunks.

\noindent$\bullet~~$\textbf{Example Retrieval}. 
In-context examples help the LLM to better follow human instruction. Instead of relying on manual specification, in-context examples can be retrieved automatically to improve the performance. 
Each example contains an optional task description, an input, and an output, which are all concatenated to form a retrieval candidate.
The user input is the concatenation of the task description and the task-specific input.
The retrieved examples and the user input are synthesized with Template~\ref{prompt:icl} to feed into the LLM.

\noindent$\bullet~~$\textbf{Tool Retrieval}. 
The LLM leverages tools to execute complex real-world tasks~\cite{qin2023toolllm,react2023yao}.
Tool retrieval efficiently provides useful tools for the LLM.
The tool's description and its API are concatenated as the retrieval candidate.
The user's request is treated as the user input. 

\subsection{Training Methodology}\label{sec:method-train}
\subsubsection{Reward Formulation}
A retrieval candidate is useful if it can facilitate the generation of the desired output (denoted as $O^*$). The absolute value of generation likelihood is not an appropriate measurement because it is prone to dramatic numerical fluctuations. Alternatively, as shown in Figure~\ref{fig:reward}, we argue that a retrieval candidate is useful if it can lead to a better ranking position of the desired output among $N$ sampled outputs from the LLM's ($\{O_i\}_{i=1}^N$). Based on this argument, we descendingly sort the sampled outputs based their generation likelihoods and compute the rank of the desired output among them when retrieval is disabled:
\begin{equation*}
    % r^a\gets|\{O_j:~&O_1,\dots,O_N\in\Theta(U)~\wedge \notag\\
    %         &p(O_j\mid U) > p(O^{gt}\mid U)\}|,
    r^a\gets \underset{O^*}{\mathrm{rank}}(\{O_1,\dots,O_N:~O_i\sim\Theta(U)\}).
\end{equation*} 
% where $\pi(X,Y)$ denotes getting the rank of $X$ among $Y$ based on the generation likelihood of the LLM.
We then compute the rank of the desired output with the same operation except that the retrieval augmentation is enabled:
\begin{equation*}
    % r^b\gets|\{O_j:~&O_1,\dots,O_N\in\Theta(\psi(U,C_i))~\wedge \notag\\
            % &p(O_j\mid \psi(U,C_i)) > p(O^{gt}\mid \psi(U,C_i))\}|.
    r^b\gets \underset{O^*}{\mathrm{rank}}(\{O_1,\dots,O_N:~O_i\sim\Theta(\psi(U,C_i))\})
\end{equation*}
Finally, the reward for the retrieval candidate $C_i$ is computed as its improvement of the rank:
\begin{equation}
    R(C_i)\gets r^a - r^b.
\end{equation}
This reward formulation is free of dealing with absolute likelihood values, but focuses on the retrieval candidate's real impact on facilitating the generation of the desired output. 

% We propose the {rank-aware reward} (Figure~\ref{fig:reward}) to estimate the usefulness of each retrieval candidate to one specific retrieval augmentation scenario.
% It first computes the rank of the desired output among $N$ sampled outputs, without conditioning on any retrieval candidate:
% \begin{align*}
%     r^a\gets|\{O_j:~&O_1,\dots,O_N\in\Theta(U)~\wedge \notag\\
%             &p(O_j\mid U) > p(O^{gt}\mid U)\}|,
% \end{align*}
% where $O^{gt}$ is the desired output.
% The metric is computed again when conditioned on a retrieval candidate $C_i$:
% \begin{align*}
%     r^b\gets|\{O_j:~&O_1,\dots,O_N\in\Theta(\psi(U,C_i))~\wedge \notag\\
%             &p(O_j\mid \psi(U,C_i)) > p(O^{gt}\mid \psi(U,C_i))\}|.
% \end{align*}
% The reward for $C_i$ is the improvement of the rank:\begin{equation}
%     R(C_i)\gets r^a - r^b.
% \end{equation}
% It is expensive to compute the reward for all retrieval candidates, thus, we only score a small subset which is mined with BGE~\cite{xiao2023bge}.

\subsubsection{Distillation Objective}\label{subsec:method-distillation}
Based on the LLM's rewards, the embedding model is learned through knowledge distillation, so that the relevance estimated by the embedder becomes consistent with the retrieval candidate's actual usefulness.
Minimizing KL-Divergence between the relevance distribution and the reward distribution is the most typical approach~\cite{shi2023replug,atlas2023izacard,yu2023augmentation}.
However, the reward distribution sometimes exhibits polarized (substantially high reward for one candidate while low for others) or flat (even reward for each candidate) patterns. 
The KL-Divergence cannot effectively distill fine-grained knowledge from these distributions.
To address this problem, we innovate a \textit{graded distillation} objective, which integrates both the absolute reward values and the relative reward orders for learning.
It consists of a series of contrastive losses, where the negatives of each loss include the lower-rewarded candidates and the in-batch candidates.
All contrastive losses are aggregated with normalized rewards as weights.
Formally, given the retrieval candidates $\{C_i\}_{i=1}^M$, their normalized rewards $w(C_i)\gets \mathrm{softmax}(R(C_*))[i]$, the objective is formulated as:
\begin{gather}
    \label{eq:loss}
    \mathcal{N}(C_i)\gets \{C:~R(C)< R(C_{i})\}\cup \mathrm{InBatch}(C_i),\notag\\
    \min\sum_{C_i}-w(C_i)\log\frac{e^{\mathrm{cos}(\boldsymbol{U},\boldsymbol{C_i})}}{\sum_{C'\in\mathcal{N}(C_i)}e^{\mathrm{cos}(\boldsymbol{U}, \boldsymbol{C'})}}.
\end{gather}
The graded distillation objective enjoys two advantages. 
On one hand, it can robustly optimize the embedder from various reward distributions. 
For the polarized rewards, it will become the one-hot contrastive learning.
For the flat rewards, it will always supervise the embedder to prioritize the more useful candidates against the less useful ones, regardless of the absolute value of the reward.
On the other hand, it incorporates in-batch negatives in the training process, which further improves the discrimination capability of the embedder. 

\subsubsection{Multi-Task Learning}
LLM-Embedder learns to support the four retrieval augmentation needs with a single model through multi-task learning.
Different retrieval tasks call for distinct semantic relationships, which may conflict with each other.
Therefore, it's important to distinguish these tasks and harmonize the their learning process. 
In this place, we tailor the multi-task learning framework with three techniques. 

\noindent$\bullet~~$\textbf{Self-Paced Learning Scheduling.}
% 每个task本身学习难度可能不一样，因此导致了模型学习每个task的速度可能不一样
% 为了防止有的简单task被过度优化，而困难task没有被彻底优化，一个常见的做法是dynamically update the task weight during training
% 然而，LLM-Embedder的多任务loss全部是同一个格式，
The intrinsic learning difficulty of each task may vary, potentially leading to differences in the model's learning pace for each task.
This may result in the over-optimization of simpler tasks and the under-optimization of more challenging tasks. 
Inspired by~\cite{loss_balanced_task_weighting2019liu}, we propose to dynamically adjust the learning pace of each retrieval task to address this problem.
% Unlike typical multi-task learning, our retrieval losses all have the same form (Equation~\ref{eq:loss}), and only one retrieval task will be optimized per step (due to the homogeneous in-batch negatives).
% Since we use Adam~\cite{adam2015kingma} optimizer, it is useless to directly scale the loss.
Specifically, we deem the loss of each retrieval task as a proxy to the learning condition of that task.
Based on it, we amplify the learning rate for lossy tasks and reduce the learning rate for already learned tasks. 
To achieve this goal, we periodically checkpoint the loss of retrieval task $T$ during training, denoted as $L^T_0$. Given the basic learning rate $\alpha$, and the current loss of the retrieval task $T$, the learning rate of the current optimization step is set to $\alpha\times\sqrt{\frac{L^T}{L^T_0}}$.

\noindent$\bullet~~$\textbf{Homogeneous Batching.}\label{subsec:method-negative}
The embedding model's discrimination capability benefits from the quality and quantity of negative samples~\cite{izacard2021unsupervised,wang2022text}, which consist of hard negatives and in-batch negatives.
% It calls for a large number of negative samples to guarantee the discrimination capability of the embedding model~\cite{izacard2021unsupervised,wang2022text}.
The vanilla batching strategy often packs training samples from different tasks in the same batch. These samples are irrelevant to each other and hence adversely influence the quality of in-batch negatives.
Instead, we gather the training samples from the same retrieval task to form every batch.
In this way, LLM-Embedder should discriminate the positive sample against $B\times M\times Z - 1$ negatives from the same retrieval task, where $B$ is the batch size, $M$ the candidate number, and $Z$ the GPU number.

\noindent$\bullet~~$\textbf{Diversified Prompting.}\label{subsec:method-instruction}
For retrieval task $T$, two unique instructions $I_U^T, I_C^T$ are assigned, which are prefixed to the user input and the retrieval candidate, respectively. 
The concatenated sequence is encoded into its embedding by LLM-Embedder: \begin{equation*}
    \boldsymbol{U}^T\gets\mathrm{encode}(I_U^T+U),~~ \boldsymbol{C}_i^T\gets\mathrm{encode}(I_C^T+C_i).
\end{equation*}
The resulting embedding $\boldsymbol{U}^T$ and $\boldsymbol{C}_i^T$ are differentiated across tasks, which helps LLM-Embedder to distinguish each task.

\section{Experiment} 
The experimental studies aim to investigate three research questions. 
\textit{RQ 1.} Can LLM-Embedder support the LLM's diverse retrieval augmentation need? ($\S$\ref{subsec:experiment-overall}) 
\textit{RQ 2.} What is LLM-Embedder's impact on each retrieval augmentation scenario? ($\S$\ref{subsec:experiment-individual})
\textit{RQ 3.} What is the individual contribution of each technique in LLM-Embedder? ($\S$\ref{subsec:experiment-ablation})

\subsection{Settings} 
\subsubsection{Training \& Evaluation}
We introduce the details of training and evaluation on the four retrieval augmentation scenarios. Statistics of all training datasets are reported in Table~\ref{tab:dataset-statistics}.
 
\noindent$\bullet~~$\textbf{Knowledge Retrieval.} 
We train LLM-Embedder with three datasets for knowledge retrieval, including MSMARCO~\cite{nguyen2016ms}, Natural Questions~\cite{kwiatkowski2019natural}, and QReCC~\cite{anantha2020open}. Note that QReCC does not have well-formed answers for generating rewards, thus, we use the annotated relevance for contrastive learning.
We include three datasets to evaluate the impact of knowledge retrieval.
1) MMLU~\cite{hendrycks2020measuring}, a multiple-choice questions dataset that covers a wide range of knowledge. We retrieve 3 passages from the MSMARCO Passage corpus~\cite{nguyen2016ms}, which are integrated as a prompt with the official Template~\ref{prompt:mmlu}. The metric is accuracy.
2) PopQA~\cite{mallen2022not}, a question answering dataset that focuses on long-tail entities. We retrieve 3 passages from Wikipedia~\cite{karpukhin2020dense}, which are integrated with the official Template~\ref{prompt:popqa}. The metric is exact match.
3) QReCC~\cite{anantha2020open}, a conversational search dataset that requires the retriever to find the relevant passage according to a conversation context. It already provides the ground-truth passage, we directly evaluate the ranking metric, i.e. NDCG@3 following previous works~\cite{LeCORE-Mao}.

\noindent$\bullet~~$\textbf{Memory Retrieval.}
We consider two tasks for memory retrieval. 
1) Long-context conversation with MSC~\cite{xu2021beyond}, where the LLM should generate the ground-truth response. 
We retrieve 1 historical dialogue turn as additional context, which is synthesized with Template~\ref{prompt:msc}. We use its training set to fine-tune LLM-Embedder.
2) Long-range language modeling with Books3~\cite{gao2020pile}, {ArXiv}~\cite{gao2020pile}, {CodeParrot}~\cite{tunstall2022natural}, and {PG19}~\cite{raecompressive2019}, where PG19 is held-out from training. We set the chunk size to 128, and maintain a recent context length of 2048. We retrieve 8 chunks and their continuation chunk to prepend to the recent context.
Perplexity is the metric for both tasks.

\noindent$\bullet~~$\textbf{Example Retrieval.}
We follow LLM-R~\cite{wang2023learning} to use in-context learning tasks from FLAN~\cite{flan2022chung} for training and evaluating the impact of example retrieval.
It consists of 9 distinct categories with 30 datasets: Closed-Book QA (CQA), Commonsense (Comm), Coreference (Coref), Paraphrase (Para), Natural Language Inference (NLI), Reading Comprehension (RC), Sentiment Analysis (Sent), Data2Text (D2T), Summarization (Summ).
We retrieve 8 examples from the union of the training set examples, which are synthesized with Template~\ref{prompt:icl}. The evaluation metric is specified in Table~\ref{tab:icl dataset}.

\noindent$\bullet~~$\textbf{Tool Retrieval.} 
We use the ToolBench~\cite{qin2023toolllm} for training and evaluating the tool retrieval performance. Akin to QReCC, this dataset does not include desired output from the LLM, hence we train LLM-Embedder with contrastive loss and directly evaluate NDCG@5. 

\subsubsection{Baselines} 
Firstly, we measure the performance of the LLM without retrieval augmentation, denoted as \textit{None}. 
Secondly, we compare with two types of retrievers. 
1) \textit{General retrievers}, which aim to support a wide range of text retrieval and representation tasks, such as question answering, entity retrieval, and duplication detection. We include the following widely-recognized baselines: BM25~\cite{robertson2009probabilistic}, Contriever~\cite{izacard2021unsupervised}, Instructor~\cite{su2022one}, RetroMAE-BEIR~\cite{liu2022retromae}, and BGE~\cite{xiao2023bge}. These methods are empirically competitive according to BEIR~\cite{thakur2021beir} and MTEB~\cite{muennighoff2022mteb} benchmarks. 
2) \textit{Task-specific embedding models}. These models are optimized for one specific retrieval augmentation scenario. We include the following baselines that excel in their respective scenario: ARR~\cite{yu2023augmentation} for knowledge retrieval, LLM-R~\cite{wang2023learning} for example retrieval, and API-Retriever~\cite{qin2023toolllm} for tool retrieval.
Since retrieval augmentation introduces additional context to the LLM, we add a simple yet strong baseline called \textit{Recency} for memory retrieval. It directly extends the context window by the length of retrieved context. 

\subsubsection{Implementation} 
We use Llama-2-7B-Chat~\cite{touvron2023llama} as the backbone LLM. 
Besides, we utilize BGE base~\cite{xiao2023bge} to initialize LLM-Embedder and fine-tune it as described in $\S$\ref{sec:method-train}. The hyper parameters during fine-tuning are shown in Table~\ref{tab:hyper parameters}.
Although using rewards from Llama-2 7B Chat, LLM-Embedder is also applicable to other LLMs and its advantage remains. The experimental results are shown in Appedix~\ref{appendix:other-llm}.
We use Flat index from faiss~\cite{Faiss} for searching.

\begin{table*}[t]
    \small
    \centering
    \begin{tabular}{p{0.20\textwidth}|cccc|c|c|c}
    \toprule
    & \multicolumn{5}{c|}{\textbf{MMLU}} & \multicolumn{1}{c}{\textbf{PopQA}} & \multicolumn{1}{|c}{\textbf{QReCC}}\\
    \cmidrule(lr){1-1}
    \cmidrule(lr){2-6}
    \cmidrule(lr){7-7}
    \cmidrule(lr){8-8}
     \textbf{Method} & 
     STEM & Social & Human & Other & All Avg. & PopQA & QReCC\\
     \midrule
     None & 0.347 & 0.533 & 0.509 & 0.497 & 0.460 & 0.206 & -- \\
     BM25 & 0.376 & 0.538 & 0.505 & 0.509 & 0.472 & 0.349 & 0.434 \\
     Instructor & 0.370 & 0.541 & 0.511 & 0.508 & 0.472 & 0.353 & 0.286 \\
     Contriever & 0.368 & 0.538 & 0.508 & 0.501 & 0.468 & 0.328 & 0.356 \\
     RetroMAE-BEIR & \textbf{0.386} & 0.546 & 0.522 & 0.528 & \underline{0.485} & 0.436 & 0.404 \\
     BGE$^*$ & \underline{0.385} & \underline{0.556} & \underline{0.519} & \textbf{0.539} & \textbf{0.490} & 0.449 & 0.386 \\
     AAR$^{\dagger}$& 0.380 & 0.550 & 0.513 & 0.529 & 0.483 & \underline{0.479} & 0.288 \\
     API-Retriever & 0.354 & 0.534 & 0.500 & 0.507 & 0.463 & 0.249 & 0.114 \\
     LLM-R & 0.363 & 0.528 & 0.502 & 0.498 & 0.463 & 0.251 & 0.023 \\
     % Conv-ANCE$^\dagger$ & -- & -- & -- & -- & -- & -- & \underline{0.456} \\
     \midrule
     LLM-Embedder (Ours)& \underline{0.385} & \textbf{0.557} & \textbf{0.523} & \underline{0.536} & \textbf{0.490} & \textbf{0.505} & \textbf{0.505} \\
     \bottomrule
    \end{tabular}
    \vspace{-6pt}
    \caption{The impact of knowledge retrieval. ``$*$'' and ``$\dagger$'' indicate the SOTA general embedder and the task-specific embedder, respectively. The best metrics are in bold, and the second-best metrics are underlined.}
    \vspace{-6pt}
    \label{tab:knowledge}
\end{table*}

\begin{table*}[t]
    \small
    \centering
    \begin{tabular}{p{0.25\textwidth} | c | ccc | c }
    \toprule
    & \multicolumn{1}{c|}{\textbf{Conversation}} &
    \multicolumn{4}{c}{\textbf{Language Modeling}}\\
    \cmidrule(lr){1-1}
    \cmidrule(lr){2-2}
    \cmidrule(lr){3-6}
     \textbf{Method} & 
      MSC & Books3 & Arxiv & CodeParrot & PG19 (o.d.) \\ 
    \midrule
     None & 19.350 & 8.819 & 3.765 & 2.766 & 10.251 \\
     Recency & \underline{13.957} & 8.739 & 3.416 & 2.599 & 10.222 \\
     BM25 & 14.651 & 8.658 & 3.311 & 2.459 & 10.196 \\
     Instructor & 14.880 & 8.662 & 3.355 & 2.476 & 10.201 \\
     Contriever & 14.213 & 8.646 & \underline{3.271} & \underline{2.444} & 10.162 \\
     RetroMAE-BEIR & 14.399 & 8.638 & 3.290 & 2.459 & 10.173 \\
     BGE$^*$ & 14.294 & \underline{8.631} & 3.291 & 2.458 & \underline{10.154} \\
     AAR & 14.700 & 8.638 & 3.326 & 2.467 & 10.181 \\
     API-Retriever & 14.783 & 8.672 & 3.386 & 2.492 & 10.183 \\
     LLM-R & 14.475 & 8.662 & 3.364 & 2.472 & 10.202 \\
     \midrule
     LLM-Embedder (Ours)  & \textbf{13.483} & \textbf{8.608} & \textbf{3.232} & \textbf{2.430} & \textbf{10.118} \\
    \bottomrule
    \end{tabular}
    \vspace{-6pt}
    \caption{The impact of memory retrieval. Recency is to directly extend the context without retrieval.} 
    \vspace{-10pt}
    \label{tab:memory}
\end{table*}

\begin{table*}[t]
    \small
    \centering
    \begin{tabular}{l | ccccccccc | c | c} 
    \toprule
     & \multicolumn{10}{c|}{\textbf{In-Context Learning}} & \textbf{Tool} \\
     \cmidrule(lr){1-1}
    \cmidrule(lr){2-11}
    \cmidrule(lr){12-12}
     \textbf{Method}
     & CQA & Comm & Coref & Para & NLI & RC & Sent & D2T & Summ & Avg & ToolBench\\
     \midrule
     None & 0.292 & 0.721 & \textbf{0.658} & 0.524 & 0.448 & 0.489 & 0.708 & 0.198 & 0.145 & 0.465 & --\\
     Random & 0.359 & 0.719 & 0.589 & 0.520 & 0.477 & 0.553 & 0.916 & 0.350 & 0.357 & 0.545 & 0 \\
     BM25 & 0.360 & 0.702 & 0.603 & 0.506 & 0.458 & 0.540 & 0.728 & 0.302 & 0.156 & 0.484 & 0.512\\
     Instructor & 0.500 & 0.777 & 0.574 & 0.631 & 0.536 & 0.622 & 0.915 & 0.460 & 0.457 & 0.604 & 0.388\\
     Contriever & 0.491 & 0.772 & 0.562 & 0.636 & 0.547 & \underline{0.630} & 0.914 & 0.438 & 0.444 & 0.601 & 0.490\\
     RetroMAE-BEIR & 0.459 & 0.774 & 0.584 & 0.576 & 0.541 & 0.603 & \textbf{0.929} & 0.466 & 0.447 & 0.594 & 0.521\\
     BGE$^*$ & 0.472 & 0.777 & 0.555 & 0.617 & 0.541 & 0.599 & \underline{0.928} & 0.472 & 0.452 & 0.597 & 0.576\\
     AAR & 0.481 & 0.780 & {0.585} & 0.589 & 0.535 & 0.604 & 0.921 & 0.445 & 0.441 & 0.594 & 0.420\\
     API-Retriever$^\dagger$ & 0.477 & 0.762 & 0.547 & 0.627 & 0.520 & 0.610 & 0.924 & \textbf{0.487} & 0.442 & 0.595 & \underline{0.802}\\
     LLM-R$^{\dagger}$ & \textbf{0.517} & \underline{0.780} & 0.583 & \textbf{0.657} & \textbf{0.615} & 0.622 & 0.906 & \underline{0.478} & \textbf{0.488} & \underline{0.626} & 0.132\\
     \midrule
     LLM-Embedder & \underline{0.516} & \textbf{0.784} & \underline{0.593} & \underline{0.656} & \underline{0.604} & \textbf{0.632} & 0.922 & 0.473 & \underline{0.474} & \textbf{0.627} & \textbf{0.865} \\
    \bottomrule
    \end{tabular}
    \vspace{-6pt}
    \caption{The impact of example retrieval and tool retrieval.}
    \label{tab:icl-tool}
    \vspace{-10pt}
\end{table*}

\begin{figure}[t]
\centering
\includegraphics[width=\linewidth]{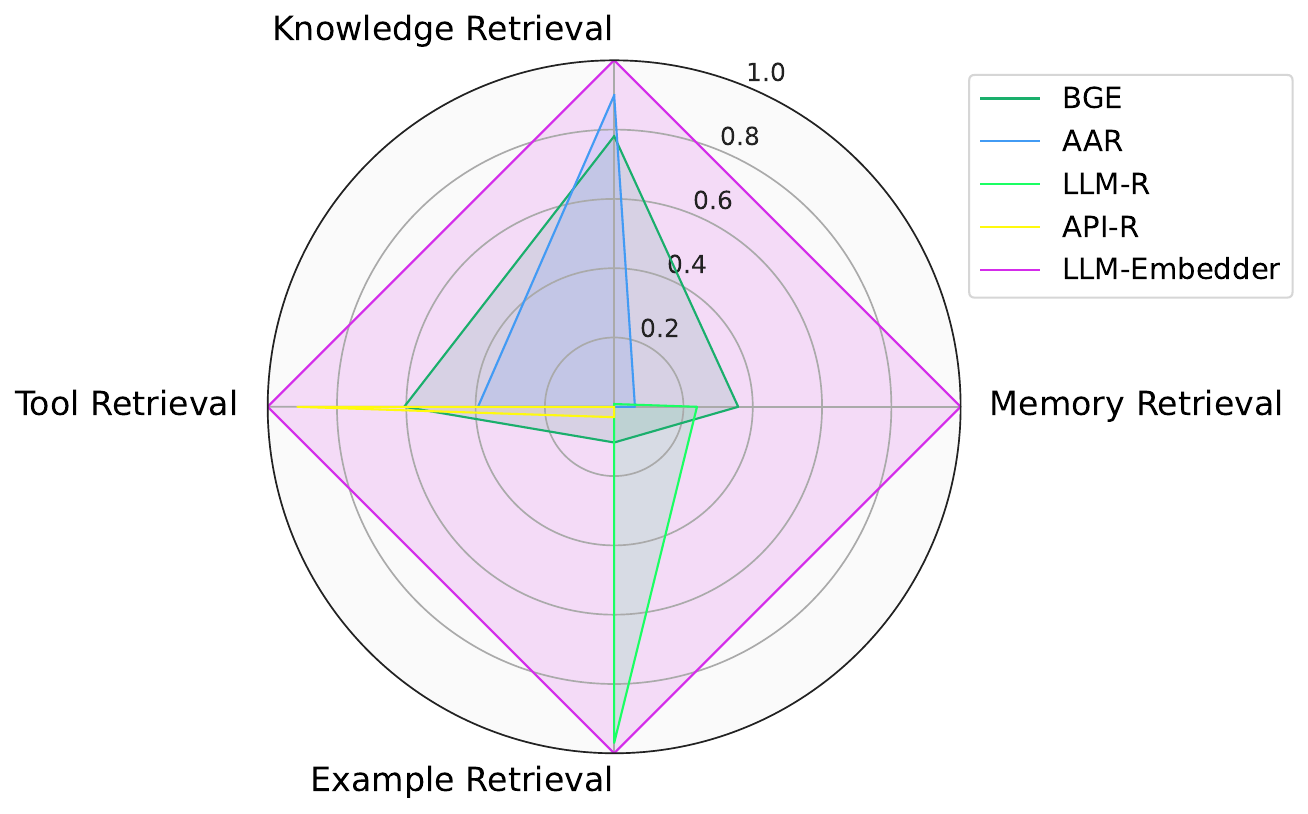}
\vspace{-20pt}
\caption{Impact of retrieval augmentation from different retrievers (metric values are min-max normalized).} 
\label{fig:radar}
\vspace{-10pt}
\end{figure}

\subsection{Overall Analysis}\label{subsec:experiment-overall}
The evaluation results of the four retrieval augmentation scenarios are presented in Table~\ref{tab:knowledge}-\ref{tab:icl-tool}.

Firstly, compared with the results without retrieval augmentation, i.e. None, LLM-Embedder delivers more precise answers with the retrieved knowledge (Table~\ref{tab:knowledge}), improved quality of long-sequence generation with the retrieved memory (Table~\ref{tab:memory}), better instruction following effect with the retrieved examples (Table~\ref{tab:icl-tool}), and more accurate tool retrieval (Table~\ref{tab:icl-tool}). 
Besides, though the LLM's performance can also by improved by other baseline retrievers, LLM-Embedder always leads to the most amplified retrieval augmentation effect across all scenarios. It outperforms all general retrievers and is competitive against task-specific retrievers, i.e. AAR for knowledge enhancement, LLM-R for example retrieval, and API-Retriever for tool retrieval. 
This observation validates that \textit{the LLM benefits from the retrieved information; meanwhile, LLM-Embedder can provide a strong and unified foundation to support diverse retrieval augmentation needs of the LLM}. 

We can also observe that the task-specific embedders optimized for one scenario result in limited performances in others, suggesting that the semantic relationships required by different retrieval scenarios are not transferable. 
To better illustrate this point, we visualize the retrieval augmentation's impact from five representative methods in Figure~\ref{fig:radar}. 
Notably, although task-specific embedders exhibit competitive performance for their targeted scenario, their impacts are severely weakened when applied on other scenarios. In contrast, LLM-Embedder demonstrates a steady and competitive performance across all scenarios. To summarize, \textit{the irrelevant or even adverse retrieval patterns can be reconciled by one unified embedding model on top of our optimized training methodology}.

\subsection{Individualized Analysis}
\label{subsec:experiment-individual}
\noindent$\bullet~~$\textbf{Knowledge Retrieval}. 
The evaluation results of knowledge retrieval are reported in Table~\ref{tab:knowledge}. We make the following observations. 
1) Benefit of external knowledge.
On both MMLU and PopQA, we can observe significant empirical advantages of the retrieval augmentation methods compared with the plain LLM, i.e. None. 
Among all retrieval methods, LLM-Embedder is able to return the most accuracy knowledge, leading to the best retrieval augmentation effect on both datasets.
2) Distinction among datasets. The impact of knowledge retrieval is more noticeable on PopQA than MMLU. This is because PopQA is more knowledge-intensive, with a focus on questions about long-tail entities. 
Moreover, the baseline embedding models fail to handle conversational search queries, resulting in their inferior NDCG compared with BM25 on QReCC. In contrast, LLM-Embedder significantly outperfoms all baselines on QReCC, again verifying its versatility.

\noindent$\bullet~~$\textbf{Memory Retrieval}. 
The evaluation results of memory retrieval are reported in Table~\ref{tab:memory}.
On one hand, baseline retrievers underperform the Recency baseline on MSC, which translates to the negative impact of the retrieved conversation compared with the recent one. This observation underscores the challenges in effective memory retrieval. 
On the other hand, the LLM-Embedder retains its superior performance, reducing the perplexity against the all baseline methods on all datasets. 

\noindent$\bullet~~$\textbf{Example Retrieval}. 
The evaluation results of example retrieval are reported in Table~\ref{tab:icl-tool}. We have the following observations. 
1) Compared with random examples, using retrieved examples yields improved performances in most cases. This finding underscores the effect of example retrieval for helping the LLM to properly follow instructions. 
2) BM25's performance is substantially weaker than its performance in other scenarios. This discrepancy can be attributed to the specific nature of in-context learning, where useful examples may have low lexical similarity with the user input.

\noindent$\bullet~~$\textbf{Tool Retrieval}. The evaluation results of example retrieval are reported in Table~\ref{tab:icl-tool}. We observe that the task-specific method, i.e. the API retriever, beats other baseline methods by a large margin. This is because these baselines are unfamiliar with tools and hence fail to properly estimate the relevance. However, LLM-Embedder continues to maintain the leading position, highlighting its unfied support for diverse retrieval tasks. 

\subsection{Ablation Studies}\label{subsec:experiment-ablation}
\begin{table}[tb]
    \scriptsize
    \centering 
    \begin{tabular}{l@{ }| cccc}
    \toprule
    \textbf{Method} & \textbf{Knwl.} & \textbf{Mem.} & \textbf{Expl.} & \textbf{Tool} \\
    \midrule
    LLM-Embedder & \textbf{0.505} & \textbf{13.483} & \textbf{0.627} & \textbf{0.865} \\ %\midrule
    \quad w.o. Rank-Aware Reward & 0.485 & 14.253 & 0.622 & 0.861 \\
    \quad w.o. Graded Distillation & 0.492 & 13.547 & 0.610 & 0.854 \\
    \quad w.o. Self-Paced Scheduling & 0.492 & 13.883 & 0.619 & 0.809 \\    
    \quad w.o. Homogeneous Batching & 0.447 & 14.183 & 0.605 & 0.836 \\
    \quad w.o. Diversified Instruction & 0.503 & 13.942 & 0.619 & 0.828 \\ 
    
    \bottomrule
    \end{tabular}
    \vspace{-6pt}
    \caption{Ablation studies of LLM-Embedder.}
    % Each technical design in LLM-Embedder's training recipe contributes to its effectiveness and versatility.} 
    \vspace{-10pt}
    \label{tab:ablation}
\end{table}

The ablation studies are performed to to evaluate the impact from each technical factor. The evaluation results are reported in Table~\ref{tab:ablation}. 

% The ablation studies aim to evaluate the contribution of each technical design in LLM-Embedder's training methodology.
% including the rank-aware reward, the graded distillation, and the three-dimension multi-task learning framework (diversified instruction, homogeneous in-batch negative, loss-balanced learning rate scaling). 
% The results are reported in Table~\ref{tab:ablation}. 

For ``\textit{w.o. Rank-Aware Reward}'', we switch to the typical likelihood-based reward formulation \cite{shi2023replug}. 
Notably, the performance on knowledge retrieval and memory retrieval substantially decreases. 
We conjecture that in both scenarios, the generation likelihood of the desired output drastically fluctuate, resulting in the inaccurate measurement of the retrieval candidate's usefulness.

% For ``\textit{w.o. Rank-Aware Reward}'', we replace the rank-aware reward with the likelihood reward.
% Notably, the performance on knowledge retrieval and memory retrieval substantially decreases. 
% We conjecture that in both scenarios, the generation likelihood of the desired output drastically fluctuate, resulting in the inaccurate measurement of the retrieval candidate's usefulness.

For ``\textit{w.o. Graded Distillation}'', the graded distillation objective is replaced by the typical KL-divergence \cite{atlas2023izacard}. As introduced, graded distillation can stay robust to the polarized or flat rewards, which leads to more effective usage of the LLM's feedback. In this place, we can observed that LLM-Embedder's performance is reduced when graded distillation is disabled, especially for example retrieval.

% robustly optimize the embedder from the polarized or flat reward distribution so that the distillation becomes more effetive. It can be observed that LLM-Embedder's performance is reduced once this technique is removed, especially for example retrieval.

For ``\textit{w.o. Self-Paced Scheduling}'', the learning rate is the kept static for all retrieval tasks during fine-tuning. 
We can observe that the performance of tool retrieval drops significantly. This is because the learning for this scenario does not proceed at the same pace as other scenarios, necessitating the dynamic control over learning speed for different retrieval tasks.

For ``\textit{w.o. Homogeneous Negatives}'', the homogeneous in-batch negatives are disabled. This change reduces the discrimination capability of the embedder, because a great portion of the in-batch negative samples will come from different tasks, which are irrelevant to the target one. As we can observe, LLM-Embedder's performance is decreased due to such a change, especially for knowledge retrieval, where LLM-Embedder should discriminate the relevant passage from a massive corpus. 

For ``\textit{w.o. Diversified Instruction}'', we remove the task-specific instructions in fine-tuning and evaluation. Without this technique, it becomes harder for the embedding model to distinguish different retrieval tasks. This intuition is consistent with the observed result, as LLM-Embedder's performance decreases across all tasks.

\section{Conclusion}
In this work, we present LLM-Embedder, a unified embedding model to support the LLM's diverse retrieval augmentation needs, including knowledge retrieval, memory retrieval, example retrieval, and tool retrieval. 
We propose three key techniques to facilitate the training of LLM-Embedder, spanning from reward formulation, distillation objective, and multi-task learning recipe. Our experiments show LLM-Embedder's empirical advantages over both general and task-specific embedding models across all evaluation scenarios. This highlights its effectiveness as a foundational building block to support the retrieval augmentation of the LLM.

\section{Limitations}
A few recent studies incorporate large language models as the embedding backbone and achieve new state-of-the-art performance.
However, LLM-Embedder is a BERT-base scale model. Its scaling effect remains unexplored.
Besides, LLM-Embedder is specifically tailored for the four retrieval scenarios. For tasks that fall outside its scope of coverage, such as documentation retrieval, the effectiveness of the LLM-Embedder may not be as robust as that of a strong general embedding model like BGE.

\section{Ethical Considerations}
LLM-Embedder is an embedding model that maps the text into high-dimensional vectors and relies on vector similarity to determine relevance between texts. Therefore, it inherits the potential risks of the embedding model family. Specifically, LLM-Embedder may process a large amount of personal or sensitive data, which must be handled with consent. There is also the security concern as recent works have proven it possible to decrypt the original textual information from embedded vectors. Lastly, it may perpetuate and amplify biases present in the training data, leading to unfair or discriminatory outcomes.

\section*{Acknowledgement}
This research is supported by National Science and Technology Major Project(2023ZD0121504).

\bibliographystyle{acl_natbib}
\bibliography{custom}

\appendix

\section{Prompt Templates}
\begin{prompt}[title={Prompt \thetcbcounter: Rank-Aware Reward (Knowledge)}, label=prompt:knowledge]
Knowledge:\\
<Passage>\\
\\
Q: <Question> A:\\
\end{prompt}

\begin{prompt}[title={Prompt \thetcbcounter: MMLU}, label=prompt:mmlu]
Knowledge:\\
<Passage 1>\\
<Passage 2>\\
<Passage 3>\\
\\
The following are multiple-choice questions (with answers) about <subject>.\\
\\
<Question>\\
A. <Option 1>\\
B. <Option 2>\\
C. <Option 3>\\
D. <Option 4>\\
Answer:
\end{prompt}

\begin{prompt}[title={Prompt \thetcbcounter: PopQA}, label=prompt:popqa]
Knowledge:\\
<Passage 1>\\
<Passage 2>\\
<Passage 3>\\
\\
Q: <Question 1> A: <Answer 1>\\
Q: <Question 2> A: <Answer 2>\\
$\dots$\\
Q: <Question 15> A: <Answer 15>\\
Q: <Question> A:
\end{prompt}

\begin{prompt}[title={Prompt \thetcbcounter: Multi-Session Chat}, label=prompt:msc]
Speaker 1: <Retrieved/Recent Utterance 1>\\
Speaker 2: <Retrieved/Recent Utterance 2>\\
Speaker 1: <Utterance 1>\\
Speaker 2:
\end{prompt}

\begin{prompt}[title={Prompt \thetcbcounter: In-Context Learning}, label=prompt:icl]
<Example 1 Input><Example 1 Ouptut>\\
\\
<Example 2 Input><Example 2 Ouptut>\\
$\dots$\\
<Example 8 Input><Example 8 Ouptut>\\
\\
<Input>
\end{prompt}

\section{Dataset Details}
\label{appendix:dataset}
The detailed information of in-context learning datasets is reported in Table~\ref{tab:icl dataset}. The statistics of all training and evaluation datasets are reported in Table~\ref{tab:dataset-statistics}. The average lengths of long-range language modeling datasets are reported in Table~\ref{tab:lrlm dataset}.

\begin{table}[tb]
    \centering
    \begin{tabular}{lc}
    \toprule
       \textbf{Dataset} & \textbf{Average Length} \\
       \midrule
       Books3 & 101010\\
       Arxiv & 26735\\
       CodeParrot & 217364 \\
       PG19 & 90447 \\
    \bottomrule
    \end{tabular}
    \caption{Average lengths of long-range language modeling datasets.}
    \label{tab:lrlm dataset}
\end{table}

\begin{sidewaystable*}[tb]
\footnotesize
\centering
\begin{tabular}{|c|c|c|c|c|c|}
\hline
\textbf{Dataset name} & \textbf{Category} & \textbf{\#Train Sample} & \textbf{\#Test Sample} & \textbf{Metric} & \textbf{Evaluation Strategy}\\
\hline
ARC Challenge~\cite{ARC} & Close QA & 1,117 & 1,165 & Accuracy & Likelihood\\
\hline
ARC Easy~\cite{ARC} & Close QA & 2,241 & 2,365 & Accuracy & Likelihood\\
\hline
NQ~\cite{kwiatkowski2019natural} & Close QA & 87,925 & 3,610 & Exact Match & Generation\\
\hline
COPA~\cite{COPA} & Commonsense & 400 & 100 & Accuracy & Likelihood\\
\hline
HellaSwag~\cite{HellaSwag} & Commonsense & 39,905 & 10,042 & Accuracy & Likelihood\\
\hline
PIQA~\cite{PIQA} & Commonsense & 16,113 (held out) & 1,838 & Accuracy & Likelihood\\
\hline
Winogrande~\cite{Winogrande} & Coreference & 40,398 & 1,267 & Accuracy & Likelihood\\
\hline
WSC~\cite{WSC} & Coreference & 554 & 104 & Accuracy & Likelihood\\
\hline
WSC273~\cite{WSC} & Coreference & 0 (held out) & 273 & Accuracy & Likelihood\\
\hline
CommonGen~\cite{CommonGen} & Data-to-text & 67,389 & 4,018 & ROUGE-L & Generation\\
\hline
DART~\cite{DART} & Data-to-text & 62,659 & 2,768 & ROUGE-L & Generation\\
\hline
E2E NLG~\cite{E2E} & Data-to-text & 33,525 & 1,847 & ROUGE-L & Generation\\
\hline
MNLI (m)~\cite{MNLI} & NLI & 392,702 & 9,815 & Accuracy & Likelihood\\
\hline
MNLI (mm)~\cite{MNLI} & NLI & 392,702 & 9,832 & Accuracy & Likelihood\\
\hline
RTE~\cite{RTE} & NLI & 2,490 & 277 & Accuracy & Likelihood\\
\hline
SNLI~\cite{SNLI} & NLI & 549,367 & 9,824 & Accuracy & Likelihood\\
\hline
QNLI~\cite{QNLI} & NLI & 104,743 (held out) & 5,463 & Accuracy & Likelihood\\
\hline
MRPC~\cite{MRPC} & Paraphrase & 3,668 & 408 & Accuracy & Likelihood\\
\hline
PAWS~\cite{PAWS} & Paraphrase & 49,401 & 8,000 & Accuracy & Likelihood\\
\hline
QQP~\cite{QQP} & Paraphrase & 363,846 & 40,430 & Accuracy & Likelihood\\
\hline
BoolQ~\cite{BoolQ} & Reading Comp. & 9,427 & 3,270 & Accuracy & Likelihood\\
\hline
MultiRC~\cite{MultiRC} & Reading Comp. & 27,243 & 4,848 & F1 & Likelihood\\
\hline
OpenBook QA~\cite{OpenBookQA} & Reading Comp. & 4,957 & 500 & Accuracy & Likelihood\\
\hline
SQuAD v1~\cite{SQuADv1} & Reading Comp. & 87,599 & 10,570 & Exact Match & Generation\\
\hline
Sentiment140~\cite{Sentiment140} & Sentiment & 1,600,000 & 359 & Accuracy & Likelihood\\
\hline
SST2~\cite{SST2} & Sentiment & 67,349 & 872 & Accuracy & Likelihood\\
\hline
Yelp~\cite{Yelp} & Sentiment & 490,456 (held out) & 33,285 & Accuracy & Likelihood\\
\hline
AESLC~\cite{AESLC} & Summarize & 13,181 & 1,750 & ROUGE-L & Generation\\
\hline
AGNews~\cite{AGNews} & Summarize & 120,000 & 7,600 & Accuracy & Likelihood\\
\hline
Gigaword~\cite{Gigaword} & Summarize & 2,044,465 & 730 & ROUGE-L & Generation\\
\hline
Total & n.a. & 6.3M & 177k & n.a. & n.a.\\
\hline
Total (sampled) & n.a. & 591k & 177k & n.a. & n.a.\\
\hline
\end{tabular}%
\caption{Detailed information of in-context learning datasets.}
\label{tab:icl dataset}
\end{sidewaystable*}

\begin{table*}[tb]
    \centering
    \begin{tabular}{l|l|c|c|c}
        \toprule
        \textbf{Scenario} & \textbf{Dataset} & \textbf{Corpus Size} & \textbf{\#Training Samples} & \textbf{\#Testing Samples}\\
        \midrule
        \multirow{4}{*}{Knowledge Retrieval} & MSMARCO & 8841823 & 400870 & -- \\
        & NQ & 21051324 & 58622 & --\\
        & MMLU & 8841823 & -- & 14042 \\
        & PopQA & 21051324 & -- & 14267 \\
        & QReCC & 54573064 & 29596 & 8209 \\
        \midrule
        \multirow{5}{*}{Memory Retrieval} & MSC & -- & 48925 & 2763 \\
        & Books3 & -- & 10000 & 1000 \\
        & Arxiv & -- & 10000 & 757 \\
        & CodeParrot & -- & 10000 & 1000 \\
        & PG19 & -- & -- & 1000 \\
        \midrule
        Example Retrieval & Misc. & 6283120 & 591359 & 177230 \\
        \midrule
        Tool Retrieval & ToolBench & 10439 & 87322 & 100 \\
        \midrule
        Total & -- & -- & 1333911 & -- \\
        \bottomrule
    \end{tabular}
    \caption{Statistics of all training and evaluation datasets.}
    \label{tab:dataset-statistics}
\end{table*}

\section{Implementation Details}

\begin{table*}[tb]
    \centering
    \begin{tabular}{l|l|c|C{5.5cm}}
        \toprule
        \textbf{Scenario} & \textbf{Task} & \textbf{Input} & \textbf{Instruction}\\
        \midrule
        \multirow{7}{4cm}{Knowledge Retrieval} & \multirow{3}{4cm}{Conversational Search} & Query & Encode this query and context for searching relevant passages: \\
        \cmidrule{3-4}
        & & Key & Encode this passage for retrieval:\\
        \cmidrule{2-4}
        & \multirow{4}{*}{Others} & Query & Represent this query for retrieving relevant documents: \\
        \cmidrule{3-4}
        & & Key & Represent this document for retrieval:\\
        \midrule
        \multirow{8}{*}{Memory Retrieval} & \multirow{4}{4cm}{Long-Context Conversation} & Query & Embed this dialogue to find useful historical dialogues: \\
        \cmidrule{3-4}
        & & Key & Embed this historical dialogue for retrieval: \\
        \cmidrule{2-4}
        & \multirow{4}{4cm}{Long-Range Language Modeling} & Query & Embed this text chunk for finding useful historical chunks: \\
        \cmidrule{3-4}
        & & Key & Embed this historical text chunk for retrieval: \\
        \midrule
        \multirow{4}{*}{Example Retrieval} & \multirow{4}{4cm}{In-Context Learning} & Query & Convert this example into a vector to look for useful examples: \\
        \cmidrule{3-4}
        & & Key & Convert this example into vector for retrieval: \\
        \midrule
        \multirow{4}{*}{Tool Retrieval} & \multirow{4}{4cm}{Tool Retrieval} & Query & Transform this user request for fetching helpful tool descriptions: \\
        \cmidrule{3-4}
        & & Key & Transform this tool description for retrieval: \\
        \bottomrule
    \end{tabular}
    \caption{Instructions for each task.}
    \label{tab:instruction}
\end{table*}

\begin{table*}[tb]
    \centering
    \begin{tabular}{l|c}
        \toprule
        \#GPU & 8$\times$A100 (40G) \\
        \#Hard Negative ($M$) & 7\\
        \#Sampled Outputs ($N$) & 10 \\
        Batch Size Per GPU ($B$) & 100 \\
        Optimizer & AdamW \\
        Learning Rate ($\alpha$) & 5e-5 \\
        Learning Rate Checkpoint Step & 1000 \\
        Weight Decay & 0.01 \\
        Scheduler & Linear with warm-up of 0.2 \\
        Max Steps & 10000 \\
        Gradient Checkpointing & \ding{51} \\
        \bottomrule
    \end{tabular}
    \caption{Hyper parameter settings for fine-tuning.}
    \label{tab:hyper parameters}
\end{table*}

\begin{table*}[t]
    \centering
    \begin{tabular}{l|l|ccccc}
        \toprule
        \textbf{LLM} & \textbf{Embedder} & \textbf{MMLU} & \textbf{PopQA} & \textbf{ICL} & \textbf{MSC} & \textbf{Arxiv}\\
        \midrule
        \multirow{3}{*}{Llama-2-7B-Chat} & None & 0.460 & 0.206 & 0.465 & 19.350 & 3.765 \\
        & BGE & \textbf{0.490} & 0.449 & 0.597 & 14.294 & 3.291 \\
        & LLM-Embedder & \textbf{0.490} & \textbf{0.505} & \textbf{0.627} & \textbf{13.483} & \textbf{3.232}\\
        \midrule
        \multirow{3}{*}{Aquila-7B-Chat} & None & 0.450 & 0.203 & 0.515 & 16.011 & 3.120\\
        & BGE & 0.483 & 0.398 & 0.573 & 14.184 & 2.791\\
        & LLM-Embedder & \textbf{0.485} & \textbf{0.440} & \textbf{0.590} & \textbf{14.184} & \textbf{2.735}\\
        \midrule
        \multirow{3}{*}{Qwen-7B-Chat} & None & 0.556 & 0.239 & 0.535 & 21.047 & 2.789\\
        & BGE & \textbf{0.579} & 0.445 & 0.633 & 16.206 & 2.517\\
        & LLM-Embedder & 0.576 & \textbf{0.478} & \textbf{0.646} & \textbf{15.452} & \textbf{2.482}\\
        \midrule
        \multirow{3}{*}{Baichuan2-7B-Chat} & None & 0.523 & 0.236 & 0.491 & 18.971 & 2.751\\
        & BGE & \textbf{0.553} & 0.441 & 0.596 & 16.076 & 2.444\\
        & LLM-Embedder & 0.551 & \textbf{0.485} & \textbf{0.618} & \textbf{15.589} & \textbf{2.413}\\
        \midrule
        \multirow{3}{*}{Llama-2-13B-Chat} & None & 0.539 & 0.289 & 0.461 & 14.733 & 3.236\\
        & BGE & \textbf{0.560} & 0.460 & 0.620 & 11.688 & 2.904\\
        & LLM-Embedder & 0.558 & \textbf{0.503} & \textbf{0.644} & \textbf{11.538} & \textbf{2.854}\\
        \bottomrule
    \end{tabular}
    \caption{The impact of LLM-Embedder on different LLMs.}
    \label{tab:generalization}
\end{table*}

\subsection{Instructions}
The instructions used for each retrieval task are shown in Table~\ref{tab:instruction}.

\subsection{Training Settings}
The hyper parameter settings for training LLM-Embedder are reported in Table~\ref{tab:hyper parameters}.

\section{Impact of LLM-Embedder on Different LLMs}\label{appendix:other-llm}
We evaluate the impact of LLM-Embedder when augmenting different LLMs to validate its generalization ability. 
Specifically, we utilize Aquila-7B-Chat~\cite{Aquila}, Qwen-7B-Chat~\cite{Qwen}, Baichuan2-7B-Chat~\cite{Baichuan2}, and Llama-2-13B-Chat~\cite{touvron2023llama}. The results are shown in Table~\ref{tab:generalization}. 
We report the average accuracy for MMLU, accuracy for PopQA, the average score for in-context learning, and perplexity for both Multi-Session Chat and Arxiv. Note that we do not replicate the evaluation of tool learning and conversational search because their performances are directly measured by retrieval metrics. 

We can observe that our conclusions in Section~\ref{subsec:experiment-overall} still hold. First of all, retrieval from the external world benefits LLM's performance in all four scenarios, since the performance of the plain LLM (i.e. None) underperforms retrieval-augmented one (BGE and LLM-Embedder). 
Besides, our proposed LLM-Embedder is able to generalize well to other LLMs and maintain its superiority over BGE on most datasets (PopQA and ICL in particular). This observation highlights the practical effectiveness and versatility of LLM-Embedder.

\end{document}